\documentclass[pdflatex,sn-mathphys-num]{sn-jnl}

\usepackage{graphicx}%
\usepackage{multirow}%
\usepackage{amsmath,amssymb,amsfonts}%
\usepackage{amsthm}%
\usepackage{mathrsfs}%
\usepackage[title]{appendix}%
\usepackage{xcolor}%
\usepackage{textcomp}%
\usepackage{manyfoot}%
\usepackage{booktabs}%
\usepackage{algorithm}%
\usepackage{algorithmicx}%
\usepackage{algpseudocode}%
\usepackage{listings}%
\usepackage{tikz, pgfplots}
\usetikzlibrary{shapes.geometric, arrows, positioning}
\usepackage{csvsimple}
\pgfplotsset{compat=1.18}
\usepackage{array}
\usepackage[utf8]{inputenc}
\usepackage{pdflscape}
\usepackage{rotating}
\usepackage{longtable}
\usepackage{threeparttablex}
\usepackage{booktabs}
\usepackage{caption}

\theoremstyle{thmstyleone}%
\theoremstyle{thmstyletwo}%

\theoremstyle{thmstylethree}%

\begin{document}

\title[Article Title]{Systematic Review of Pituitary Gland and Pituitary Adenoma Automatic Segmentation Techniques in Magnetic Resonance Imaging}

\author*[1, 2, 3]{\fnm{Mubaraq} \sur{Yakubu}}\email{mubaraq.yakubu@kcl.ac.uk}

\author[4]{\fnm{Navodini} \sur{Wijithilake}}\email{navodini.wijethilake@kcl.ac.uk}

\author[4]{\fnm{Jonathan} \sur{Shapey}}\email{Jonathan.Shapey@kcl.ac.uk}

\author[4]{\fnm{Andrew} \sur{King}}\email{andrew.king@kcl.ac.uk}

\author[1, 5, 6]{\fnm{Alexander} \sur{Hammers}}\email{alexander.hammers@kcl.ac.uk}

\affil*[1]{\orgdiv{King's College London \& Guy's and St Thomas' PET Centre}, \orgname{King's College London}, \orgaddress{\street{Westminster Bridge Rd}, \city{London}, \postcode{SE1 7EH}, \state{Greater London}, \country{United Kingdom}}}

\affil[2]{\orgdiv{Radiology Department}, \orgname{Aminu Kano Teaching Hospital}, \orgaddress{\street{Zaria Rd}, \city{Kano}, \postcode{700233}, \state{Kano}, \country{Nigeria}}}

\affil[3]{\orgdiv{Medical Artificial Intelligence (MAI) Lab}, \orgname{Crestview Radiology-NOHIL CT and MRI Centre}, \orgaddress{\street{Igbobi}, \city{Ikeja-Main Land}, \postcode{100252}, \state{Lagos}, \country{Nigeria}}}

\affil[4]{\orgdiv{Biomedical Engineering and Imaging Science}, \orgname{King's College London}, \orgaddress{\street{Lambeth Palace Rd}, \city{London}, \postcode{SE1 7EU}, \state{Greater London}, \country{United Kingdom}}}

\affil[5]{\orgdiv{Research Department of Biomedical Computing, School of Biomedical Engineering and Imaging Sciences}, \orgname{King's College London}, \orgaddress{\street{Westminster Bridge Rd}, \city{London}, \postcode{SE1 7EH}, \state{Greater London}, \country{UK}}}

\affil[6]{\orgdiv{Research Department of Early Life Imaging, School of Biomedical Engineering and Imaging Sciences}, \orgname{King's College London}, \orgaddress{\street{Westminster Bridge Rd}, \city{London}, \postcode{SE1 7EH}, \state{Greater London}, \country{UK}}}

\abstract{\textbf{Purpose}: Accurate segmentation of both the pituitary gland and adenomas from magnetic resonance imaging (MRI) is essential for diagnosis and treatment of pituitary adenomas. This systematic review evaluates automatic segmentation methods for improving the accuracy and efficiency of MRI-based segmentation of pituitary adenomas and the gland itself. \textbf{Methods}: We reviewed 34 studies that employed automatic and semi-automatic segmentation methods. We extracted and synthesized data on segmentation techniques and performance metrics (such as Dice overlap scores). \textbf{Results}: The majority of reviewed studies utilized deep learning approaches, with U-Net-based models being the most prevalent. Automatic methods yielded Dice scores of 0.19--89.00\% for pituitary gland and 4.60--96.41\% for adenoma segmentation. Semi-automatic methods reported 80.00--92.10\% for pituitary gland and 75.90--88.36\% for adenoma segmentation.

\textbf{Conclusion}: Most studies did not report important metrics such as MR field strength, age and adenoma size. Automated segmentation techniques such as U-Net-based models show promise, especially for adenoma segmentation, but further improvements are needed to achieve consistently good performance in small structures like the normal pituitary gland.  Continued innovation and larger, diverse datasets are likely critical to enhancing clinical applicability.}

\keywords{Pituitary Adenoma, Pituitary Gland, Magnetic Resonance Imaging (MRI), Automatic Segmentation, Deep Learning, UNet, Semi-Automatic Segmentation}

\maketitle

\section{Introduction}\label{sec1}

 Pituitary adenomas (PAs), also known as as pituitary neuroendocrine tumours \cite{tsukamoto2023imaging}, are heterogeneous slow-growing brain tumours caused by abnormal growth of the pituitary gland (PG).  PAs are common (14\% in autopsy studies) \cite{ezzat2004prevalence}, but almost all are benign \cite{molitch2017diagnosis}. Functional PAs, often microadenomas ($<$1\,cm), may cause debilitating hormonal disorders such as Cushing’s disease~\cite{castle2023mri}, while larger, non-functional macroadenomas ($>$1\,cm) and giant PAs ($>$4\,cm) can exert mass effects, leading to headaches and visual disturbances by compressing structures like the optic chiasm and cavernous sinus~\cite{chaichana2013role,yildiz1999radiotherapy,molitch2009pituitary,davies2016assessing}.
\\
Clinical investigation of PAs involves the use of magnetic resonance imaging (MRI) of the pituitary region \cite{yao2017metabolic}. Common MRI sequences for imaging PAs include T1-weighted, T2-weighted, and Fluid Attenuated Inverted Recovery (FLAIR) \cite{chatain2017potential}. Contrast medium, such as gadolinium-DTPA, is often used, with a standard dose of 0.1 mmol/kg, to improve delineation of PAs and surrounding structures, particularly in small functional PAs \cite{buchfelder2010modern, niendorf1987dose}. T1-weighted (with and without contrast) and T2-weighted MRI sequences provide satisfactory visualization of pituitary adenomas in most cases \cite{bashari2019modern}. There have been reported differences in the MRI appearance of different types and stages of PAs in MRI images acquired by different sequences and pre/post contrast  \cite{ntali2018epidemiology}. However, unlike other brain conditions, PA suffers from a lack of publicly accessible data to support research and model development. A notable exception is the Cheng 2015 dataset \cite{cheng2017brain}, also referred to as the Figshare dataset \cite{cheng2015enhanced}.
\\
Several techniques have been employed to segment anatomical structures. Traditional methods such as thresholding, region growing, region merging and splitting, clustering, edge detection, atlasing and model-based approaches have laid the foundation for automated or semi-automated image segmentation \cite{pham2000current}. The rise of deep learning techniques has revolutionized the automatic segmentation of medical images, especially for brain tumors like PAs \cite{ramesh2021review}. A recent review highlighted the prominence of U-Net \cite{UNET2015}, a deep learning-based fully convolutional network, and its variants demonstrating superior performance in brain tumor segmentation tasks \cite{wangetalreview2023brainsystematic}. This shift towards deep learning-based techniques is driven by the need for more accurate, precise and efficient approaches to the diagnosis and treatment planning of PAs \cite{havaei2017brain}.
\\
We found significant variation in the methodological approaches, model types, and reporting of findings across studies related to the automatic segmentation of normal PG and PA in MRI. This review explores and systematically integrates these varied perspectives comprehensively, providing a broader understanding of the current state of the field and explored how accurate automated segmentation algorithms for PG and PA are.
The main aim of this work is to systematically review the available literature on automatic segmentation of PG and PA that used magnetic resonance images.

\section{Methods}
The systematic review was conducted in
accordance with PRISMA guidelines \cite{page2022prisma}.
\subsection{Eligibility Criteria}
Articles that focus on improved PG or PA segmentation using MRI to enable accurate diagnosis are the main focus of this systematic review. All articles that met the following inclusion criteria were reviewed:
\begin{enumerate}
    \item Full length peer-reviewed articles on pituitary imaging segmentation.
    \item Description of an automated machine learning or deep learning algorithm for the segmentation of MRI images specifically targeting the pituitary or sellar region.
\end{enumerate}
We excluded case reports, literature reviews, conference abstracts, book chapters, meta-analyses, and editorials as well as articles that did not mention the use of manual and automatic segmentation or MR acquisition protocol in patients with PA or PG imaging.
The information sources including search strategy and other data collection processes are available in the supplementary material.

\section{Results}
\subsection{Study Selection}
\subsubsection{Search and Selection Process}

\begin{figure}[h]
\centering

\tikzstyle{startstop} = [rectangle, rounded corners, minimum width=3cm, minimum height=1cm,text centered, text width=2.5cm, draw=black, fill=green!30]
\tikzstyle{io} = [trapezium, trapezium left angle=70, trapezium right angle=110, minimum width=3cm, minimum height=1cm, text centered, text width=2.5cm, draw=black, fill=blue!30]
\tikzstyle{process} = [rectangle, minimum width=3.5cm, minimum height=1cm, text centered, text width=3.5cm, draw=black, fill=orange!30]
\tikzstyle{decision} = [diamond, minimum width=3cm, minimum height=1cm, text centered, text width=3cm, draw=black, fill=pink!30]
\tikzstyle{arrow} = [thick,->,>=stealth]

\begin{tikzpicture}[node distance=2cm]
\node (start) [startstop] {Data identified from Scopus (n=302), PubMed (n=130), Web of Science (n=285)};
\node (promb1a) [process, right of=start, xshift=2cm] {Duplicate records removed before screening (n=460)};
\node (RHstart) [decision, right of=promb1a, xshift=3cm] {Additional records identified from websites (n=0), citation searching (n=1)};
\node (in1) [io, below of=start, yshift=-0.5cm] {Records screened (n=353)};
\node (promb2a) [process, right of=in1, xshift=2cm] {Records excluded (n=261)};
\node (pro1) [io, below of=in1] {Reports sorted for retrieval (n=92)};
\node (pro2) [process, right of=pro1, xshift=2cm] {Reports not retrievable (n=3)};
\node (dec1) [io, below of=pro1, yshift=-0.5cm] {Reports assessed for eligibility (n=89)};
\node (pro2b) [process, right of=dec1, xshift=2cm] {Reports excluded: No segmentation outcome (n=47), no overlap metrics (n=3), similar approach (n=3), book chapter (n=2), wrong formulas (n=1) };
\node (pro2ba) [io, right of=pro2, xshift=3cm] {Reports sought for retrieval (n=1)};
\node (pro3ba) [io, below of=pro2ba, yshift=-0.5cm] {Reports assessed for eligibility (n=1)};
\node (out1) [io, below of=dec1] {Total studies (n=33)};
\node (stop) [startstop, below of=out1] {34 studies included};

\draw [arrow] (start) -- (in1);
\draw [arrow] (start) -- (promb1a);
\draw [arrow] (in1) -- (promb2a);
\draw [arrow] (pro1) -- (pro2);
\draw [arrow] (in1) -- (pro1);
\draw [arrow] (pro1) -- (dec1);
\draw [arrow] (RHstart) -- (pro2ba);
\draw [arrow] (pro2ba) -- (pro3ba);
\draw [arrow] (dec1) -- (pro2b);
\draw [arrow] (dec1) -- (out1);
\draw [arrow] (pro3ba) |- (stop);
\draw [arrow] (out1) -- (stop);
\end{tikzpicture}
\caption{Flowchart of the search and selection process for study inclusion}
\label{fig:search_selection_process}
\end{figure}
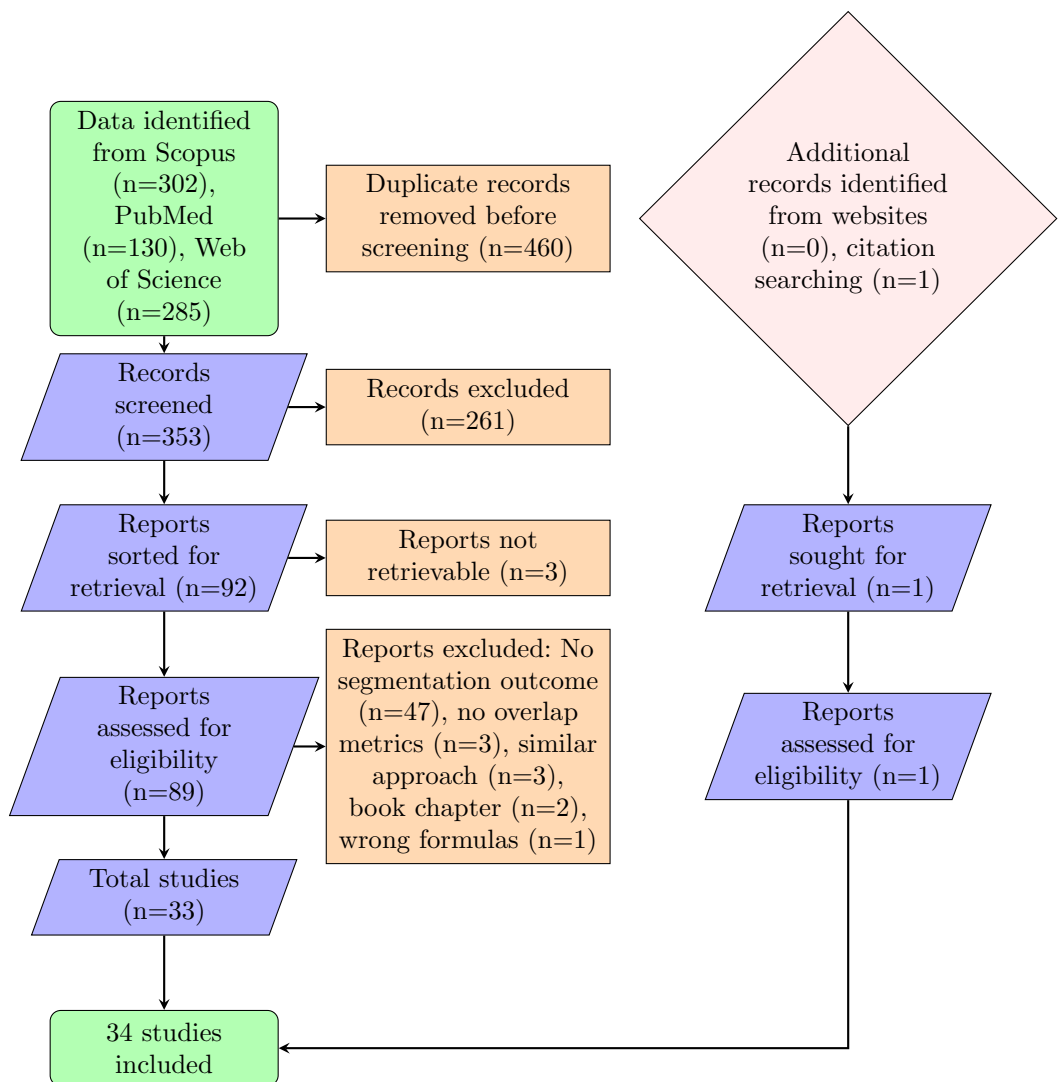

\begin{figure}[h]
\centering

\tikzstyle{startstop} = [rectangle, rounded corners, minimum width=3cm, minimum height=1cm,text centered, text width=2.5cm, draw=black, fill=green!30]
\tikzstyle{process} = [rectangle, minimum width=3cm, minimum height=1cm, text centered, text width=3cm, draw=black, fill=orange!30]
\tikzstyle{decision} = [diamond, minimum width=2cm, minimum height=0.5cm, text centered, text width=3cm, draw=black, fill=pink!30]
\tikzstyle{arrow} = [thick,->,>=stealth]

\begin{tikzpicture}[node distance=2cm]

\node (start) [startstop] {Records identified};
\node (abstract) [process, below of=start] {Title/Abstract screening};
\node (rayyan) [process, below of=abstract] {Screened in Rayyan \cite{ouzzani2016rayyan}};
\node (criteria) [decision, below of=rayyan, yshift=-1cm] {Meets inclusion criteria?};
\node (conflict) [process, right of=criteria, xshift=5cm] {Disagreements?};
\node (resolve) [process, below of=conflict] {Resolve via third reviewer};
\node (fulltext) [startstop, below of=criteria, yshift=-3cm] {Full text included};

\draw [arrow] (start) -- (abstract);
\draw [arrow] (abstract) -- (rayyan);
\draw [arrow] (rayyan) -- (criteria);
\draw [arrow] (criteria) -- (fulltext);
\draw [arrow] (criteria) -- (conflict);
\draw [arrow] (conflict) -- (resolve);
\draw [arrow] (resolve) |- (fulltext);

\end{tikzpicture}
\caption{Selection process for inclusion of studies}
\label{fig:selection_flowchart}
\end{figure}
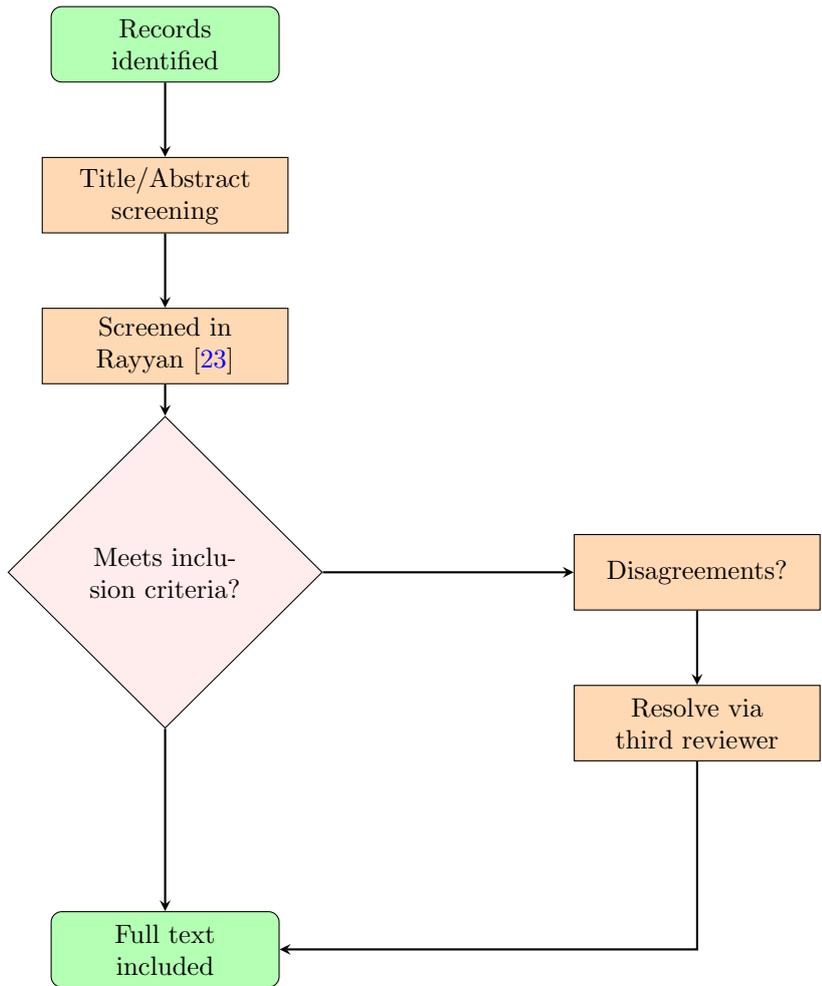

The detailed processes are summarized in Figures \ref{fig:search_selection_process} and \ref{fig:selection_flowchart}.

\clearpage

\clearpage

\subsubsection{Segmentation Approaches for Pituitary Gland MRI Studies}

\begin{figure}[H]
    \centering
    \includegraphics[width=\textwidth]{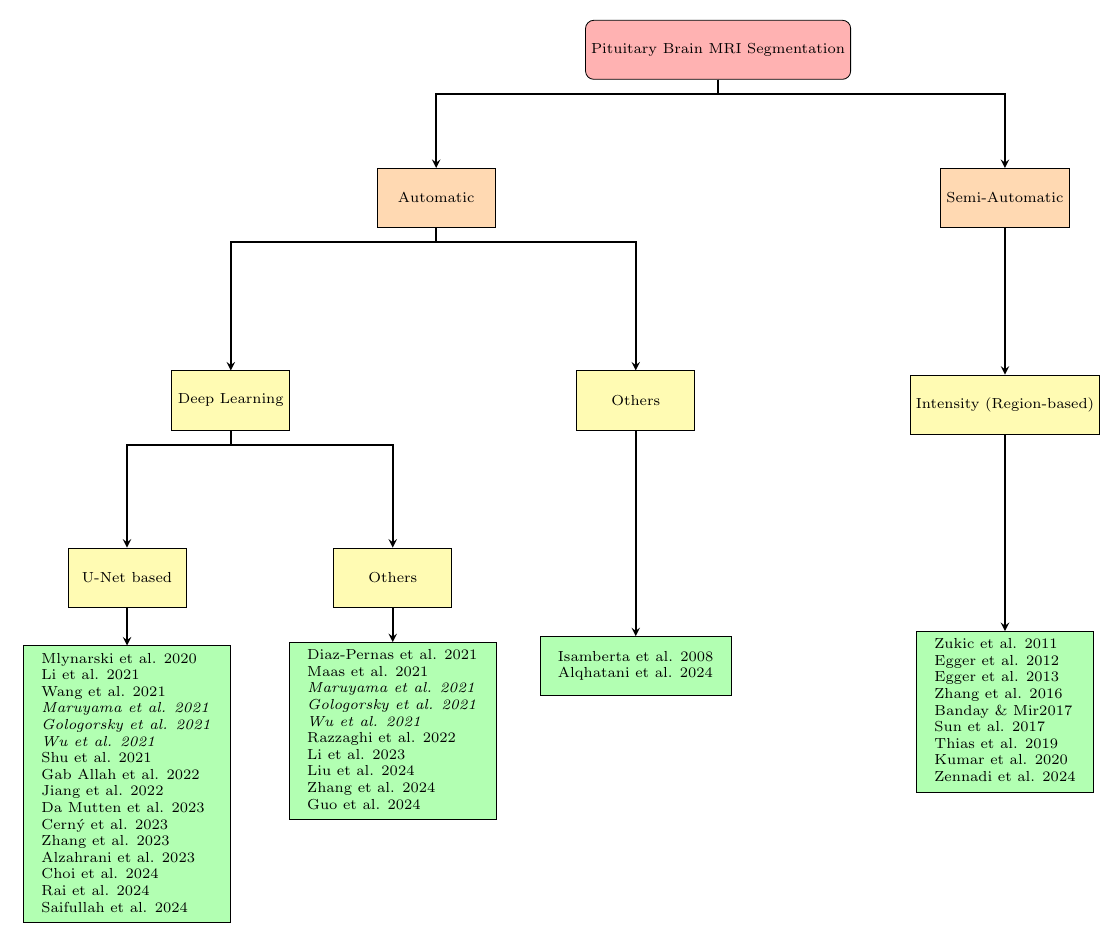}
    \caption{Segmentation approaches for pituitary MRI Studies. Studies in \textit{italics} utilized both U-Net and other Deep Learning methods, i.e. are listed twice}
    \label{fig:segmentation_approaches}
\end{figure}

To categorize the segmentation techniques used in the selected studies, we classified them based on the level of automation and the methods employed. The segmentation strategies are shown in Figure \ref{fig:segmentation_approaches}.

\clearpage

\subsection{Study Characteristics}

A total of 34 studies were included in this review, each focusing on different segmentation approaches and MRI modalities, characterized as follows:

\textbf{Segmentation Approaches:} 13 studies employed purely \textbf{UNet-based models} for segmentation, three used a combination of \textbf{UNet-based} and other deep learning models, eight explored \textbf{other deep learning models}, two applied an automatic segmentation approach without \textbf{deep learning}, and eight utilized \textbf{semi-automatic approaches}.

\textbf{MRI Field Strength:} six studies used 1.5T MRI scans, two studies used 3T MRI scans, five studies used both 1.5T and 3T scans, one study used 1T, 1.5T and 3T, and the remaining 20 studies did not specify the MRI field strength, including nine of the reviewed studies that utilized the \textit{Cheng 2015 dataset} \cite{cheng2017brain}.

\subsubsection{Cheng Dataset Description}
The \textit{Cheng 2015 dataset} \cite{cheng2017brain} provides T1-weighted contrast-enhanced MRI scans of the pituitary region. The full dataset consists of 3064 slices from 233 patients and includes three types of brain tumors: 930 PA, 708 meningioma, and 1426 glioma slices. The images were acquired between September 2005 and October 2010 at Nanfang Hospital (Guangzhou, China) and the General Hospital of Tianjin Medical University (Tianjin, China). Each scan has an in-plane resolution of 512 × 512 pixels, with a pixel spacing of 0.49 × 0.49 mm\textsuperscript{2}, a slice thickness of 6 mm, and an inter-slice gap of 1 mm. A gadolinium-based contrast agent (Gd-DTPA) was administered at a standard dose of 0.1 mmol/kg at an injection rate of 2 ml/s.

\subsection{Results of Individual Studies}

The results are categorized based on the type of segmentation models used and further subdivided based on their application to PG and PA. Performance metrics, such as Dice Similarity Coefficient and other relevant findings, are summarized in Tables \ref{tab:Deep_Learning_Pituitary}, \ref{tab:Deep_Learning_Adenoma}, \ref{tab:others_Pituitary}, and \ref{tab:Others_Adenoma}. Studies included by inclusion criteria but which were later excluded can be found in Appendix \ref{secA1}.

\clearpage
\begin{sidewaystable}[ht]
\centering
\scriptsize
\caption{Outcome of Automatic PG Segmentation Studies}
\label{tab:Deep_Learning_Pituitary}
\begin{tabular}{p{15mm} p{15mm} p{15mm} p{20mm} p{15mm} p{15mm} p{15mm} p{20mm} >{\centering\arraybackslash}p{20mm} >{\raggedleft\arraybackslash}p{10mm}} 
\toprule
\textbf{Study} & \textbf{Dataset Used} & \textbf{MRI Strength (Tesla)} & \textbf{MS Soft} & \textbf{Ground Truth (Images)} & \textbf{Models Used} & \textbf{Model Number} & \textbf{Model Description} & \textbf{2D vs. 3D} & \textbf{Dice} \\
\midrule
\cite{isambertetal2008evaluation} & Local & 1.5 & N/A & 11 & 1 & 1 & ABAS & 2D & 30.00 \\
 & & & & & & & & & \\
\cite{mlynarskietal2020anatomically} & Local & N/A & N/A & 29 & 1 & 1 & Modified 2D U-Net (OVMA) & 2D & 79.70 \\
 & & & & & & & & & \\
\cite{wangetal2021development} & Local & 3.0 & ITK-Snap & 213 & 1 & 1 & Gated-shaped U-Net & 2D & 60.00 \\
 & & & & & & & & & \\
\cite{maruyamaetal2021simultaneous} & ADNI & 1.5 \& 3.0 & MATLAB Image Labeler & 450 & 6 & 1 & AlexNet & 2D & 46.86 \\
 & & & & & & 2 & GoogleNet & & 42.39 \\
 & & & & & & 3 & ResNet50 & & 47.32 \\
 & & & & & & 4 & SegNet & & 49.96 \\
 & & & & & & 5 & SegNet with VGG16-Weighting factor & & 58.65 \\
 & & & & & & 6 & U-Net & & 0.19 \\
 & & & & & & & & & \\
\cite{gologorskyetal2022generating} & ABIDE & N/A & 3D Slicer & 333 & 7 & 1 & UNET3D & 2D & 79.10 \\
 & & & & & & 2 & VNET & & 77.10 \\
 & & & & & & 3 & CONDSEG & & 79.30 \\
 & & & & & & 4 & OBELISK 96 & & 75.20 \\
 & & & & & & 5 & OBELISK 144 & & 74.50 \\
 & & & & & & 6 & UNETR & & 72.00 \\
 & & & & & & 7 & Ensemble & & 79.60 \\
 & & & & & & & & & \\
\cite{cernyetal2023fully} & Local & 1.5 \& 3.0 & ITK-snap software & 521 & 1 & 1 & U-Net & 3D & 61.10 \\
 & & & & & & & & & \\
\cite{alzahranietal2023geometric} & Local & N/A & N/A & 32 & 3 & 1 & U-Net-MRIu & 3D & 35.00 \\
 & & & & & & 2 & U-Net-MRIeCT & & 32.00 \\
 & & & & & & 3 & U-Net-MRIeMRI & & 67.00 \\
 & & & & & & & & & \\
\cite{choi2024deeppgsegnet} & Local & 3.0 & ITK-Snap & 153 & 1 & 1 & 3D U-Net & 3D & 89.00 \\
 & & & & & & & & & \\
\cite{liu2024caln} & Local & 3.0 \& 1.5 & N/A & 195 & 1 & 1 & CALN & 2D & 84.02 \\
 & & & & & & & & & \\
\cite{guo2025computer} & Local & 3.0 \& 1.5 & ITK-Snap & 2586 & 1 & 1 & Mask R-CNN & 2D & 47.77 \\
\bottomrule
\end{tabular}
\begin{flushleft}
2D: Two-Dimensional,  
3D: Three-Dimensional,  
ABAS: Atlas-based automatic segmentation software,  
ABIDE: Autism Brain Imaging Data Exchange,  
ADNI: Alzheimer's Disease Neuroimaging Initiative,  
CALN: Channel Attention (Long-Short-Term-Memory) Network, 
CONDSEG: Conditional Segmentation,  
MS Soft: Manual Segmentation Software,  
MRI: Magnetic Resonance Imaging,  
MSR-Net: Multi-Scale Residual Network,  
N/A: Not available,  
OBELISK: Object Boundary Extraction using Learned Image Structures,  
OVMA: One Voxel Mismatch Allowed,  
PG: Pituitary Gland,  
UNET3D: 3D version of U-Net,  
UNETR: U-Net Transformer,  
VNET: V-Net
\end{flushleft}
\end{sidewaystable}

\clearpage

\begin{sidewaystable}[h]
\centering
\scriptsize
\caption{Outcome of Automatic PA Segmentation Studies}
\label{tab:Deep_Learning_Adenoma}
\begin{tabular}{p{10mm} p{15mm} p{15mm} p{15mm} p{15mm} p{10mm} p{15mm} p{15mm} p{20mm} >{\centering\arraybackslash}p{15mm} >{\raggedleft\arraybackslash}p{10mm}}
\toprule
\textbf{Study} & \textbf{Dataset Used} & \textbf{Adenoma Size} & \textbf{MRI Strength (Tesla)} & \textbf{MS Soft} & \textbf{Ground Truth (Images)} & \textbf{Models Used} & \textbf{Model Number} & \textbf{Model Description} & \textbf{2D vs. 3D} & \textbf{Dice} \\
\midrule
\cite{diaz-pernasetal2021deep} & Cheng 2015 & N/A & N/A & N/A & 930 & 1 & 1 & Multiscale CNN & 2D & 81.30 \\
 & & & & & & & & & & \\
\cite{maasetal2021quicktumornet} & Cheng 2015 & N/A & N/A & N/A & 930 & 1 & 1 & Modified QuickNAT & 2D & 81.20 \\
 & & & & & & & & & & \\
\cite{lietal2021image} & Local & Micro-PA \& Macro-PA & 1.5 \& 3.0 & MITK & 185 & 1 & 1 & Res U-Net & 3D & 80.93 \\
 & & & & & & & & & & \\
\cite{wangetal2021development} & Local & N/A & 3.0 & ITK-Snap & 213 & 1 & 1 & Gated-shaped U-Net & 2D & 89.80 \\
 & & & & & & & & & & \\
\cite{wuetal2021deep} & NTUH & N/A & N/A & N/A & 155 & 5 & 1 & V-Net\_dropout & 2D & 27.00 \\
& & & & & & & 2 & Deconvnet & & 38.00 \\
& & & & & & & 3 & U-Net & & 7.00 \\
& & & & & & & 4 & PSPNet4 & & 24.00 \\
& & & & & & & 5 & DeepMedic & & 29.00 \\
 & & & & & & & & & & \\
\cite{shuetal2021three} & Local & Micro, Macro \& Giant PA & 1.5 & ITK-Snap & 243 & 2 & 1 & U-Net-All type PA & 3D & 80.30 \\
& & & & & & & 2 & U-Net-Primary NFPA & & 85.30 \\
 & & & & & & & & & & \\
\cite{gaballahetal2023edge} & Cheng 2015 & N/A & N/A & N/A & 930 & 1 & 1 & Edge U-Net & 2D & 87.28 \\
 & & & & & & & & & & \\
\cite{jiangetal2022improved} & Local & N/A & N/A & LabelMe & 500 & 1 & 1 & Modified U-Net & 2D & 88.87 \\
 & & & & & & & & & & \\
\cite{razzaghi2022multimodal} & Cheng 2015 & NA & N/A & N/A & 930 & 1 & 1 & Multimodal CNN & 2D & 91.08 \\
 & & & & & & & & & & \\
\cite{lietal2023preoperatively} & Local & Micro, Macro \& Giant PA & N/A & ITK-Snap & 155 & 1 & 1 & cfVB-Net & 3D & 87.70 \\
 & & & & & & & & & & \\
\cite{damuttenetal2024automated} & Local & Micro, Macro \& Giant PA & 1, 1.5 \& 3.0 & N/A & 213 & 2 & 1 & U-Net-Preoperative & 2D & 62.00 \\
& & & & & & & 2 & U-Net-Postoperative & & 4.60 \\
 & & & & & & & & & & \\
 \end{tabular}
\end{sidewaystable}
\clearpage

{\begin{sidewaystable}[t]
\ContinuedFloat
\centering
\scriptsize
\caption{Outcome of Automatic PA Segmentation Studies (continued)}
\label{tab:Deep_Learning_Adenoma}
\begin{tabular}{p{10mm} p{15mm} p{15mm} p{15mm} p{15mm} p{10mm} p{15mm} p{15mm} p{20mm} >{\centering\arraybackslash}p{20mm} >{\raggedleft\arraybackslash}p{10mm}}
\toprule
\cite{cernyetal2023fully} & Local & N/A & 1.5 \& 3.0 & ITK-Snap & 521 & 1 & 1 & U-Net & 3D & 93.40 \\
 & & & & & & & & & & \\

\cite{zhang2023pdc} & Local & N/A & N/A & LabelMe & 2000 & 1 & 1 & PDC U-Net & 2D & 88.45 \\
 & & & & & & & & & & \\
\cite{rai2024two} & Cheng 2015 & N/A & N/A & N/A & 930 & 1 & 1 & EfficientNet & 2D & 96.41 \\
 & & & & & & & & & & \\
\cite{alqhtani2024improved} & Cheng 2015 & N/A & N/A & N/A & 930 & 1 & 1 & FCM & 2D & 95.80 \\
 & & & & & & & & & & \\

\cite{zhang2024msr} & Local & N/A & N/A & LabelMe & 2105 & 1 & 1 & MSR-Net & 2D & 89.34 \\
 & & & & & & & & & & \\
\cite{saifullah2024automatic} & Cheng 2015 & N/A & N/A & N/A & 930 & 1 & 1 & Modified CNN-U-Net framework & 2D & 92.05 \\
 & & & & & & & & & & \\
\cite{guo2025computer} & Local & N/A & 3.0 and 1.5 & ITK-Snap & 1617 & 1 & 1 & Mask R-CNN & 2D & 74.97 \\
\bottomrule
\end{tabular}
\begin{flushleft}
2D: Two-Dimensional,  
3D: Three-Dimensional,  
cfVB-Net: coarse-to-fine VB-Net,  
CNN: Convolutional Neural Network,
FCM: Fuzzy C-Means
Giant PA: Giant Adenoma,  
Macro-PA: Macroadenoma,  
Micro-PA: Microadenoma,  
MRI: Magnetic Resonance Imaging,  
MS Soft: Manual Segmentation Software,  
MSR-Net: Multi-Scale Residual Network,  
N/A: Not available,  
NFPA: Non-Functioning Pituitary Adenoma,  
NTUH: National Taiwan University Hospital,  
PA: Pituitary Adenoma,  
PDC U-Net: parallel dilated convolutional network,  
PSPNet4: Pyramid Scene Parsing Network version 4,  
Res U-Net: Residual U-Net
\end{flushleft}
\end{sidewaystable}}

\clearpage

\begin{sidewaystable}[h]
\centering
\scriptsize
\caption{Outcome of Semi-Automatic PG Segmentation Studies}
\label{tab:others_Pituitary}
\begin{tabular}{p{15mm} p{15mm} p{15mm} p{15mm} p{15mm} p{20mm} p{10mm} p{20mm} >{\centering\arraybackslash}p{20mm} >{\raggedleft\arraybackslash}p{10mm}}
\toprule
\textbf{Study} & \textbf{Dataset Used} & \textbf{MRI Strength (Tesla)} & \textbf{MS Soft} & \textbf{Ground Truth (Images)} & \textbf{Models Used} & \textbf{Model Number} & \textbf{Model Description} & \textbf{2D vs. 3D} & \textbf{Dice} \\
\midrule
\cite{zhangetal2016computing} & Local & N/A & N/A & 10 & 1 & 1 & MB & 2D & 87.00 \\
 & & & & & & & & & \\
\cite{bandayandmir2017enhancement} & Local & 1.5 & N/A & 10 & 1 & 1 & MB & 2D & 92.10 \\
 & & & & & & & & & \\
\cite{zennadi2024mri} & Local & 1.5 & SPM & 100 & 1 & 1 & MPA & 3D & 80.00 \\
\bottomrule
\end{tabular}
\begin{flushleft}
2D: Two-Dimensional,  
3D: Three-Dimensional,  
MB: Morphological Based,  
MPA: Maximum Probability Atlas,  
MRI: Magnetic Resonance Imaging,  
MS Soft: Manual Segmentation Software,  
N/A: Not available,  
PG: Pituitary Gland,  
SPM: Statistical Parametric Mapping
\end{flushleft}
\end{sidewaystable}

\clearpage

\begin{sidewaystable}[h]
\centering
\scriptsize
\caption{Outcome of Semi-Automatic PA Segmentation Studies}
\label{tab:Others_Adenoma}
\begin{tabular}{p{10mm} p{15mm} p{15mm} p{15mm} p{15mm} p{15mm} p{15mm} p{15mm} p{20mm} >{\centering\arraybackslash}p{20mm} >{\raggedleft\arraybackslash}p{10mm}}
\toprule
\textbf{Study} & \textbf{Dataset Used} & \textbf{Adenoma Size} & \textbf{MRI Strength (Tesla)} & \textbf{MS Soft} & \textbf{Ground Truth (Images)} & \textbf{Models Used} & \textbf{Model Number} & \textbf{Model Description} & \textbf{2D vs. 3D} & \textbf{Dice} \\
\midrule
\cite{zukicetal2011preoperative} & Local & N/A & N/A & N/A & 10 & 1 & 1 & Balloon inflation-based & 2D & 75.92 \\
 & & & & & & & & & & \\
\cite{eggeretal2012pituitary} & Local & N/A & 1.5 & MeVisLab & 10 & 1 & 1 & Grow-Cut In Slicer & 2D \& 3D & 81.97 \\
 & & & & & & & & & & \\
\cite{eggeretal2013segmentation} & Local & N/A & 1.5 & N/A & 10 & 2 & 1 & Graph-based & 2D & 77.50 \\
 &  &  &  &  &  &  & 2 & Balloon inflation-based & & 75.90 \\
 & & & & & & & & & & \\
\cite{sunetal2017random} & N/A & N/A & N/A & ITK-Snap & 23 & 1 & 1 & Random Walk \& Graph-Cut-Based Active Contour & 2D & 88.36 \\
 & & & & & & & & & & \\
\cite{thiasetal2019brain} & Cheng 2015 & N/A & N/A & N/A & 915 & 3 & 1 & SAC & 2D & 83.76 \\
 &  &  &  &  &  &  & 2 & MACWE & & 83.76 \\
 &  &  &  &  &  &  & 3 & MGAC & & 84.39 \\
  & & & & & & & & & & \\
\cite{kumaretal2020brain} & Cheng 2015 & N/A & N/A & N/A & 930 & 1 & 1 & Edge based contouring & 2D & 83.95 \\
\bottomrule
\end{tabular}
\begin{flushleft}
2D: Two-Dimensional,  
3D: Three-Dimensional,  
MACWE: Morphological Active Contour Without Edge,  
MGAC: Morphological Geodesic Active Contour,  
MRI: Magnetic Resonance Imaging,  
MS Soft: Manual Segmentation Software,  
PA: Pituitary Adenoma,  
SAC: Snake Active Contour
\end{flushleft}
\end{sidewaystable}

\clearpage

\subsection{Results of Synthesis}
\subsubsection{Summary of Study Characteristics}

\begin{figure}[htbp]
    \centering
    \includegraphics[width=\textwidth]{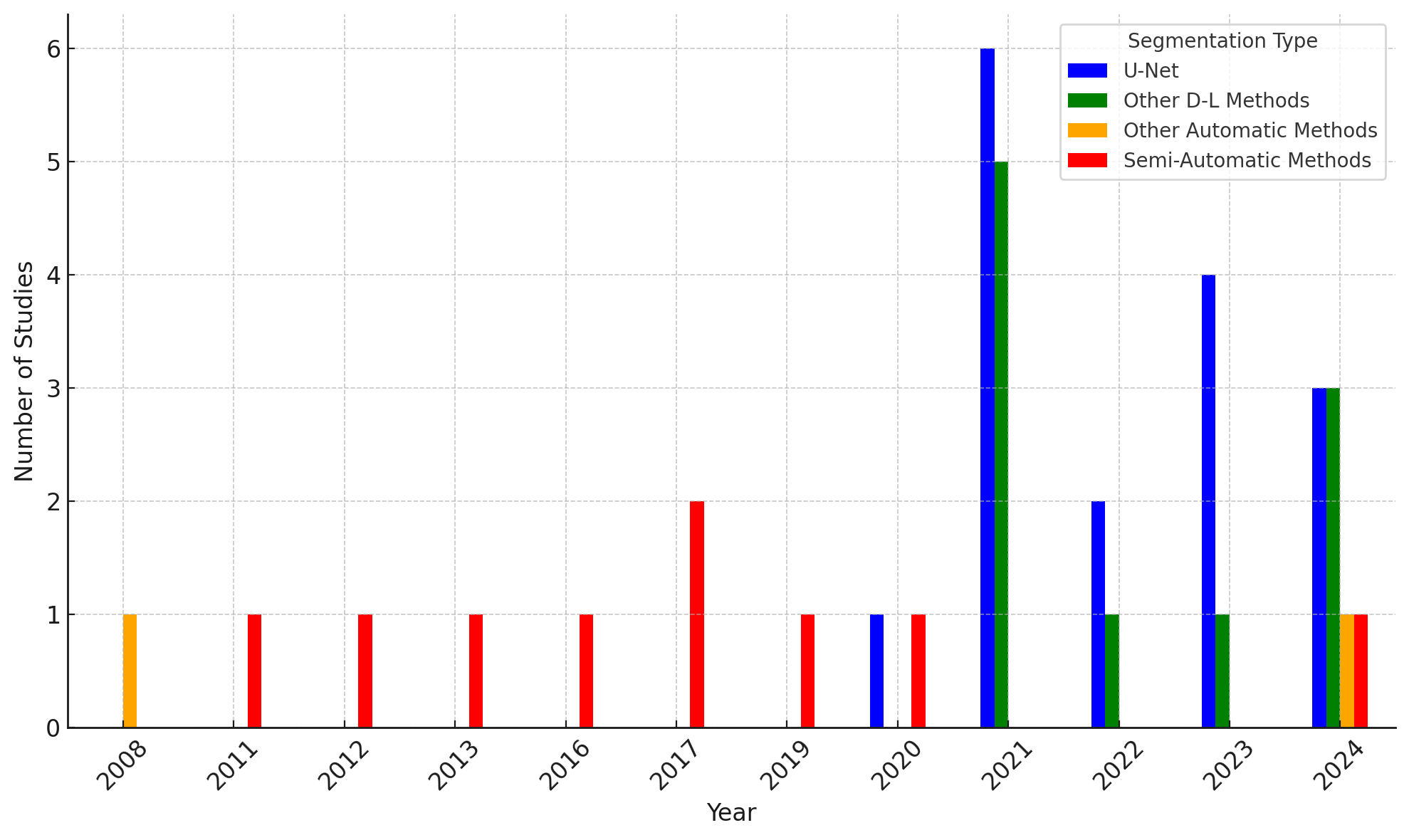}
    \caption{Number of segmentation studies per year}
    \label{fig:Number Of Segmentation Studies Per Year}
\end{figure}

The distribution of the number and type of included segmentation studies per year is shown in Figure~\ref{fig:Number Of Segmentation Studies Per Year}. The most commonly used approach was deep learning-based automatic segmentation, particularly U-Net models.

\begin{figure}[H]
    \centering
    \includegraphics[width=0.75\textwidth]{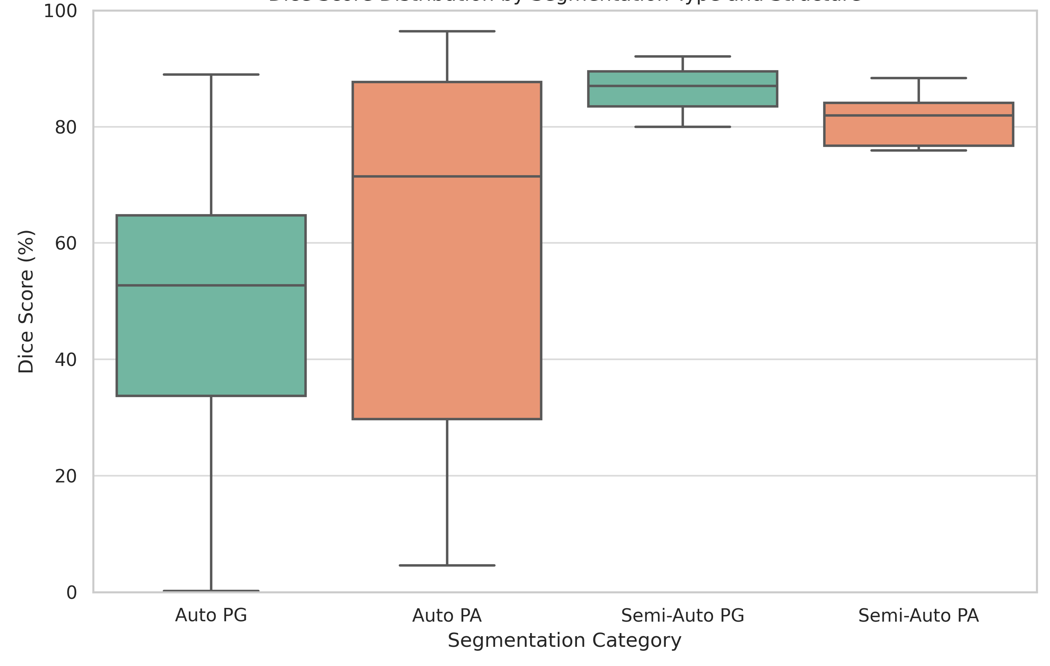}
    \caption{Dice score distributions for automated and semi-automated segmentation methods across pituitary gland (PG) and pituitary adenoma (PA) structures. Box plot horizontal lines represent the median (center), first and third quartiles (box edges), and the minimum and maximum values within 1.5× interquartile range (whiskers). No outliers were detected in this comparison.}
    \label{fig:boxplot_dice_scores_by_type}
\end{figure}

Figure~\ref{fig:boxplot_dice_scores_by_type} presents a box plot summarizing the distribution of Dice scores across four segmentation categories: automatic PG, automatic PA, semi-automatic PG, and semi-automatic PA. Automatic PG segmentation methods reported the widest performance range, from 0.19\% to 89.00\%, while one automatic PA segmentation method showed the highest overall score, reaching 96.41\%. Semi-automatic methods exhibited tighter performance distributions, with PG segmentation performance ranging from 80.00\% to 92.10\% and PA segmentation from 75.90\% to 88.36\%.

\begin{figure}[H]
    \centering
    \includegraphics[width=0.75\textwidth]{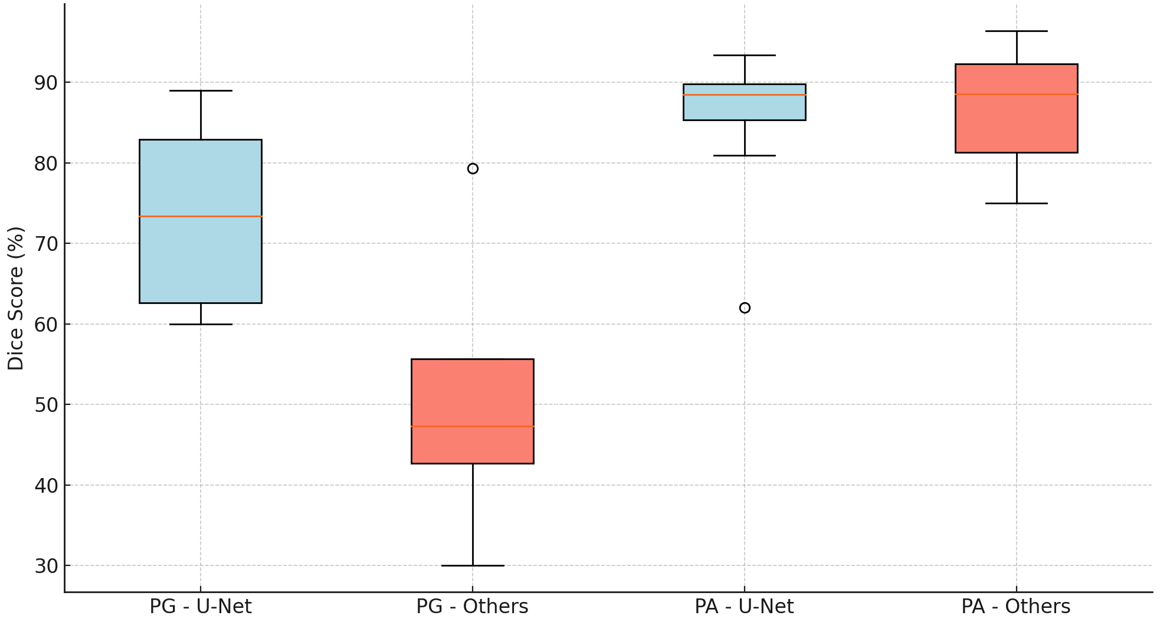}
    \caption{Box plot comparing Dice scores for U-Net-based and non-U-Net-based models across pituitary gland (PG) and pituitary adenoma (PA) segmentation. Elements are the same as Figure \ref{fig:boxplot_dice_scores_by_type}  (dots outside the whiskers [PG-Others and PA-U-Net] represent outliers). For each study, only the highest reported Dice score was used in this comparison.}
    \label{fig:pg_pa_unet_vs_others_boxplot}
\end{figure}

Figure~\ref{fig:pg_pa_unet_vs_others_boxplot} presents a comparison of Dice scores achieved by U-Net-based and non-U-Net-based models for both pituitary gland (PG) and pituitary adenoma (PA) segmentation. These comparisons are based on the highest Dice score reported by each study for the respective model category.

\begin{figure}[H]
    \centering
    \includegraphics[width=0.7\textwidth]{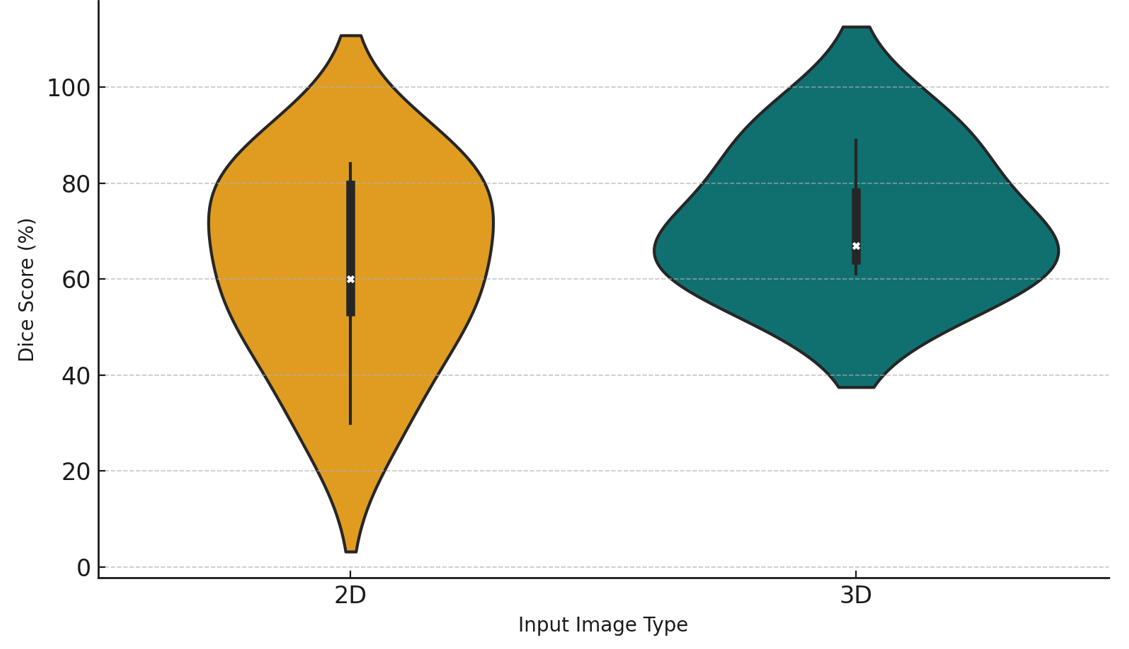}
    \caption{Violin plots of Dice score distribution for PG automatic segmentation comparing 2D and 3D input image types. The width of each violin indicates the density of data points at different Dice score values. The central white dot represents the median, the thick black bar is the interquartile range, and the thin black line shows the range within 1.5× the interquartile range. Wider sections indicate more frequent Dice scores in that range.}
    \label{fig:pg_2d_vs_3d_input_violin}
\end{figure}

 Models trained on 3D data showed a wider range of Dice scores, with the highest reported value at 89.0\%. In contrast, those trained on 2D data displayed a narrower spread, with scores ranging from 47.8\% to 84.0\%. This comparison highlights the performance variability associated with the dimensionality of the input imaging data used during model training as seen in Figure~\ref{fig:pg_2d_vs_3d_input_violin}.

\begin{figure}[H]
    \centering
    \includegraphics[width=0.7\textwidth]{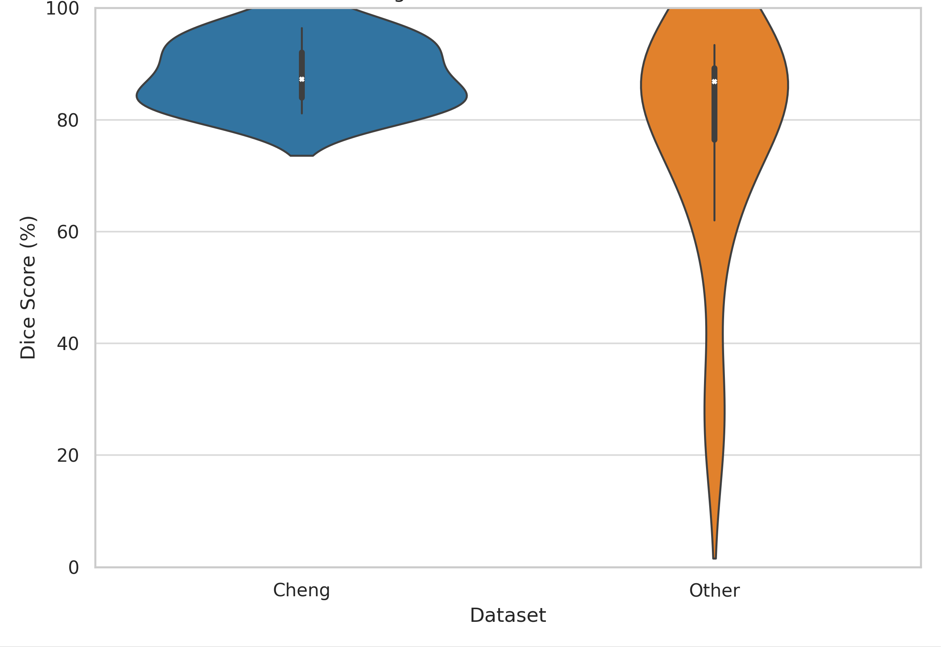}
    \caption{Violin plots of Dice score distribution for PA automatic segmentation: Cheng 2015 dataset vs other datasets. Elements are the same as Figure \ref{fig:pg_2d_vs_3d_input_violin}.}
    \label{fig:cheng_vs_other_violin}
\end{figure}

Cheng-based studies reported Dice scores within a narrower and higher range, while studies using other datasets exhibited a wider spread of Dice scores, including both lower and higher values as seen in Figure~\ref{fig:cheng_vs_other_violin}.

\begin{figure}[H]
    \centering
    \includegraphics[width=0.75\textwidth]{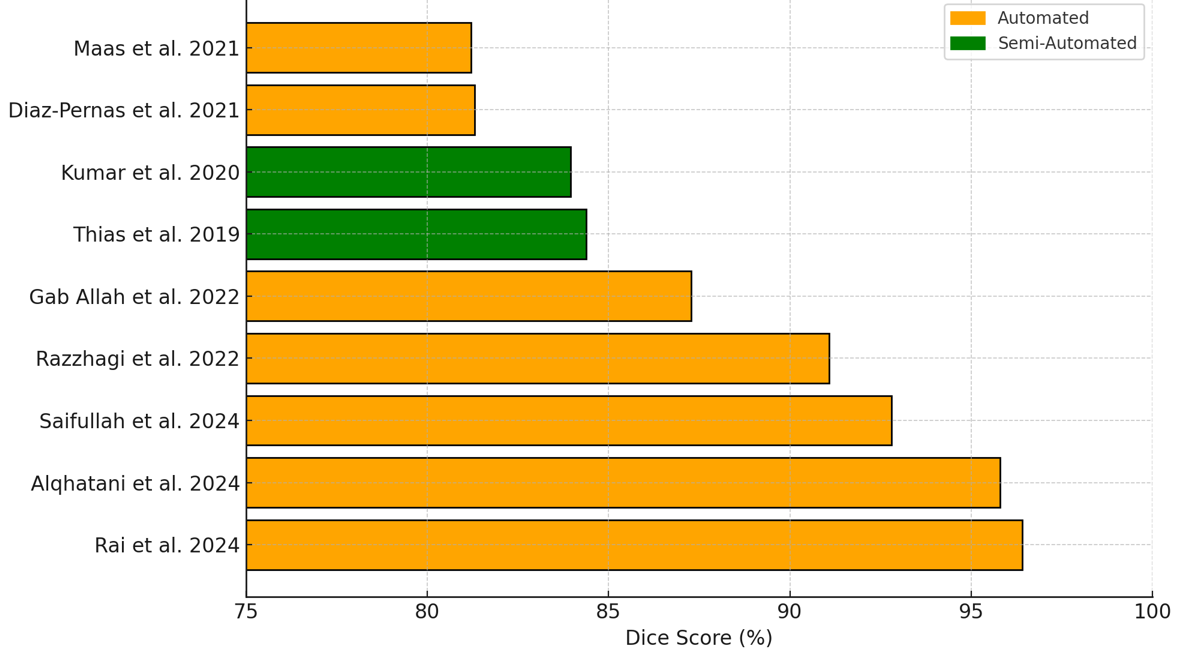}
    \caption{Dice scores of studies that utilized the Cheng 2015 dataset for pituitary adenoma segmentation, grouped by segmentation method. Automated models demonstrated a wider and generally higher performance range compared to semi-automated methods.}
    \label{fig:cheng_auto_vs_semi}
\end{figure}

Within studies that used the \textit{Cheng 2015 dataset} for PA segmentation, automatic methods reported Dice scores ranging from 81.20\% to 96.41\%, while semi-automatic methods achieved scores between 83.95\% and 84.39\%, Figure~\ref{fig:cheng_auto_vs_semi}.

Statistical comparisons were conducted using the highest Dice score reported per study. Significant differences were observed in PG segmentation between automatic and semi-automatic methods, Table~\ref{tab:statistical_comparisons}. Other comparisons, including PA segmentation methods, 2D vs. 3D input types, and dataset origin (Cheng vs. others), did not show statistically significant differences. Comparisons on field strength, age, gender, menstrual status, and pituitary or adenoma size were not possible due to insufficient reporting in the original studies.

\begin{table}[h]
\centering
\scriptsize
\caption{Statistical comparison of Dice scores across segmentation method groupings using the Mann–Whitney U (MW) and Kolmogorov–Smirnov (K–S) tests. First comparator is better.}
\label{tab:statistical_comparisons}
\begin{tabular}{p{45mm} p{15mm} p{25mm} p{15mm} p{20mm} p{25mm}}
\toprule
\textbf{Comparison} & \textbf{MW} & \textbf{MW p-value} & \textbf{K–S Statistic} & \textbf{K–S p-value} \\
\midrule
PG Auto vs Semi-Auto & 3.0 & \textbf{0.049*} & 0.800 & 0.070 \\
PA Auto vs Semi-Auto & 87.0 & 0.158 & 0.524 & 0.085 \\
U-Net vs Others (PG Auto) & 21.0 & 0.329 & 0.600 & 0.238 \\
U-Net vs Others (PA Auto) & 41.0 & 1.000 & 0.222 & 0.989 \\
Cheng vs Other (PA Auto) & 55.0 & 0.151 & 0.481 & 0.208 \\
2D vs 3D (PG Auto) & 6.0 & 0.383 & 0.571 & 0.400 \\
Auto vs Semi-Auto (Cheng) & 10.0 & 0.500 & 0.714 & 0.333 \\
\bottomrule
\end{tabular}
\begin{flushleft}
\textbf{*} Statistically significant (p\textless{}0.05)
\end{flushleft}
\end{table}

\subsubsection{Possible Causes of Heterogeneity Among Results}

The segmentation techniques used in the included studies varied, ranging from deep learning-based models such as U-Net to region-based and active contour methods. Additionally, there were notable differences in imaging modalities, with studies using 1.5T, 3T, a combination of both, and one study employing 1T, 1.5T, and 3T MRI field strengths. However, many studies did not specify the field strength used.

\subsubsection{Sensitivity Analyses Conducted}
Given the heterogeneity in segmentation techniques and the lack of relevant data reported (see supplementary material), a sensitivity analysis was not feasible or relevant.

\section{Discussion}
 
\subsection{Overview of Segmentation Techniques}

Our review identified a variety of studies focusing on the segmentation of PA and PG using MRI. These studies can be categorized into two primary segmentation approaches: automatic and semi-automatic. Within these categories, the techniques utilized varied widely with some used frequently, and some studies introduced innovative architectures and training methods to optimize performance. A single dataset with 2D slices only was very commonly used.

\subsubsection{Automatic Segmentation}

\textbf{Deep Learning Methods}

Among the 26 studies using deep learning techniques, 16 employed U-Net-based models. The performance of these models varied, highlighting the importance of dataset quality, preprocessing, and architectural modifications. Modifications such as residual blocks \cite{lietal2021image}, attention mechanisms \cite{liu2024caln}, ensemble learning \cite{gologorskyetal2022generating}, and optimization techniques \cite{saifullah2024automatic} helped enhance performance. While PG segmentation remained challenging due to anatomical complexity and small size, models like DeepPGSegNet \cite{choi2024deeppgsegnet} achieved up to 89\%, and PA segmentation with EfficientNet reached 96.41\% \cite{rai2024two}.

\textbf{Segmentation of Both PA and Residual Healthy PG}

Several studies included both PA and residual healthy PG tissue segmentation from the same subjects in their models. These studies are significant because segmenting both structures simultaneously adds complexity due to the differences in size, shape, and tissue contrast between the PA and the normal PG (or residual healthy PG), but is also more clinically significant.

- Wang et al. \cite{wangetal2021development} employed a \textit{Gated-shaped U-Net}, incorporating region-specific gating mechanisms to improve segmentation accuracy, achieving a Dice score of 89.80\% for PA segmentation and 60.00\% for the PG. The higher performance in PA segmentation was attributed to the PA's larger size and clearer boundaries compared to the smaller and more complex anatomical structure of the residual healthy PG.

- Cerny et al. \cite{cernyetal2023fully} implemented a fully automated segmentation system using a standard \textit{U-Net} architecture, achieving a high Dice score of 93.4\% for PA segmentation and 61.1\% for the residual healthy PG. Interestingly, additional pulse sequences did not enhance performance, suggesting that the core architecture was robust enough for PA segmentation without needing extra input modalities.

\textbf{Advancements and Contributions of Deep Learning Methods}

For \textit{PG segmentation}, the consistent challenge across studies has been the small size and indistinct anatomical boundaries of the pituitary gland. Nonetheless, several approaches have yielded promising results. Alzahrani et al. \cite{alzahranietal2023geometric} adopted a multi-modal approach, integrating CT and MRI contours into U-Net frameworks, which led to a Dice score of 67\%. Mlynarski et al. \cite{mlynarskietal2020anatomically} demonstrated that minor architectural modifications to the conventional U-Net, such as having one segmentation layer per class trained on three planes with majority voting, and allowing for one-voxel mismatch to assess the performance, can improve segmentation performance metrics on small structures, reporting a Dice score of 79\%. Building upon architectural enhancements, Gologorsky et al. \cite{gologorskyetal2022generating} implemented ensemble learning with a suite of 3D U-Net variants, achieving up to 79\% accuracy by leveraging diverse model outputs. Meanwhile, Liu et al. \cite{liu2024caln} introduced attention mechanisms into their architecture via the Channel Attention Long-Short-Term-Memory (CALN) model, combining Long-Short-Term-Memory (LSTM) and channel attention to enhance fine-grained spatial detail, achieving 84\%. The highest performance was observed in the study by Choi et al. \cite{choi2024deeppgsegnet}, whose \textit{DeepPGSegNet} achieved a Dice score of 89\% despite being trained on a relatively narrow age cohort, underscoring the potential of deeper, well-optimized U-Net architectures for small structure segmentation.

In contrast, \textit{PA segmentation} presents a broader structural target, especially macroadenomas and giant PAs but still benefits from the adaptability of U-Net-based models. Several studies emphasized architectural optimization to handle variations in adenoma size and morphology. Shu et al. \cite{shuetal2021three} used the dynamic, self-configuring nnU-Net framework to accommodate diverse tumor types, achieving up to 85\% Dice, with improved performance on larger lesions. Similarly, Wang et al. \cite{wangetal2021development} employed a gated-shaped U-Net incorporating spatial attention to segment both PA and residual healthy PG tissue, achieving 89\% Dice for PA. Jiang et al. \cite{jiangetal2022improved} addressed feature scaling challenges through a modified U-Net with cross-layer connections, resulting in an 88\% Dice score. Cerný et al. \cite{cernyetal2023fully} demonstrated that even a conventional 3D U-Net, when trained effectively, can achieve excellent performance—reporting a Dice score of 93\%, the highest among reviewed studies that did not utilize the Cheng 2015 dataset 2D slices \cite{cheng2017brain}. 

From a multi-scale learning perspective, \textit{Zhang et al. (2023)} \cite{zhang2023pdc} introduced parallel dilated convolutions and attention modules within their Parallel Dilated Convolutional (PDC) U-Net to enhance boundary representation, while \textit{Zhang et al. (2024)} \cite{zhang2024msr} developed Multiscale Residual Network (MSR-Net) with dual decoding paths, attaining Dice scores of 88\% and 89\%, respectively.

Other studies tackled PA segmentation through model regularization and optimization strategies. Saifullah et al. \cite{saifullah2024automatic} incorporated Particle Swarm Optimization (PSO) into a CNN-U-Net hybrid to fine-tune hyperparameters such as learning rate and dropout, yielding 92\%. \textit{Li et al. (2021)} \cite{lietal2021image} used a residual U-Net to enhance feature extraction across scales, achieving 80\%. Wu et al. \cite{wuetal2021deep}, in a comparative benchmarking study, observed that U-Net underperformed (7\%) relative to deeper architectures such as DeepMedic (29\%), illustrating the limitations of basic U-Net models when data is sparse or heterogeneous. The highest Dice score was reported by Rai et al. \cite{rai2024two}, who designed a dual-headed UNet-EfficientNet model capable of simultaneous classification and segmentation, and further improved performance through post-processing using connected component labeling, reaching 96\%.

\textbf{Other Deep Learning Approaches}

Several studies explored deep learning methods beyond the commonly used U-Net architecture, yielding diverse results in PA and PG segmentation. The multiscale feature integration demonstrated by Diaz-Pernas et al. \cite{diaz-pernasetal2021deep} and the progressive refinement strategy of cfVB-Net in \textit{Li et al. (2023)} \cite{lietal2023preoperatively} illustrate how architectural choices can directly address the challenges of segmenting small and variable anatomical structures obtaining Dice scores of 81\% and 87\% respectively. Gologorsky et al. \cite{gologorskyetal2022generating} further highlighted the benefits of ensemble learning, showing how the aggregation of complementary model outputs can yield more robust segmentation outcomes in complex regions like the sellar and para-sellar space achieving a Dice score of 79\%. 

\textbf{Comparison with UNet-Based Methods:}

\textit{Performance Metrics:} While U-Net-based architectures delivered strong results—reaching Dice scores as high as 96\% in PA segmentation (Rai et al. \cite{rai2024two}) and 89\% in PG segmentation (Choi et al. \cite{choi2024deeppgsegnet})—several non-U-Net approaches produced competitive outcomes. For example, the multiscale method by Diaz-Pernas et al. \cite{diaz-pernasetal2021deep} achieved 81\%, and the modified QuickNAT model by Maas et al. \cite{maasetal2021quicktumornet} reached 81\%. However, not all alternative architectures matched this performance; Wu et al. \cite{wuetal2021deep}, working with a limited dataset, reported Dice scores below 40\% across several tested models, underscoring the importance of data volume and quality.

\textit{Computational Requirements:} Architectures such as QuickNAT \cite{roy2019quicknat} and the multiscale CNN \cite{diaz-pernasetal2021deep} introduced greater computational demands compared to baseline U-Net models due to their pretraining and structural complexity. While U-Net’s design is known for its efficiency and ease of implementation, these advanced methods often rely on more complex layer hierarchies or ensemble strategies. Nevertheless, the added complexity may be warranted in contexts where enhanced sensitivity and specificity are crucial, particularly for variable tumor sizes or boundary detection.

\textit{Application Scenarios:} A recurring theme among these alternative approaches is their alignment with specialized use cases. \textit{Li et al. (2023)} \cite{lietal2023preoperatively} integrated segmentation with radiomics to predict the Ki67 proliferation index—a clinically relevant biomarker—highlighting the potential of tailored architectures like cfVB-Net in personalized diagnostics. Similarly, Gologorsky et al. \cite{gologorskyetal2022generating} addressed broader brain region segmentation using diverse volumetric models, reflecting a different operational focus from typical lesion-targeted pipelines.

In summary, while U-Net and its derivatives remain dominant in pituitary segmentation due to their reliability and flexibility, other deep learning strategies offer valuable alternatives in specific scenarios. These models expand the methodological toolkit for handling diverse data characteristics, enhancing performance where traditional architectures may encounter limitations.

\textbf{Other Automatic Methods}

A few studies explored non-deep learning automatic methods. Isambert et al. \cite{isambertetal2008evaluation} used an atlas-based approach (ABAS), developed from synthetic data \cite{bondiau2005atlas}, for PG segmentation and achieved a low Dice score (30\%) due to anatomical variability. In contrast, Alqhatani et al. \cite{alqhtani2024improved} applied Fuzzy C-Means (FCM) clustering with preprocessing enhancements for PA, achieving 95\%. These approaches show potential but remain highly dependent on structure type and dataset conditions.

\textbf{Comparison with Deep Learning Approaches:}

The Dice score of 30\% reported by Isambert et al. \cite{isambertetal2008evaluation} using the ABAS atlas-based method highlights the limitations of non-learning approaches in segmenting small structures like the PG, where deep learning models such as U-Net have achieved up to 79\% \cite{mlynarskietal2020anatomically} and 89\% \cite{choi2024deeppgsegnet}. While multi-atlas methods may perform comparably to deep learning in other contexts \cite{yaakub2020brain}, ABAS’s synthetic-anatomy approach performance remains low. Alqhatani et al. \cite{alqhtani2024improved} achieved a high Dice score of 95\% for PA segmentation using FCM clustering and enhancement techniques, nearly matching the 96\% by Rai et al. \cite{rai2024two} using a UNet-EfficientNet model. However, both relied on the same dataset \cite{cheng2017brain}, potentially inflating performance.

\subsubsection{Semi-Automatic Segmentation}

Intensity (region-based) segmentation approaches have proven effective in PG and PA delineation, offering an alternative to deep learning when computational resources or annotated datasets are limited. Zhang et al. \cite{zhangetal2016computing} and Banday \& Mir \cite{bandayandmir2017enhancement} combined wavelet transforms with mathematical morphology for PA segmentation, achieving Dice scores of 87\% and 92\%, respectively. For PG, Zennadi et al. \cite{zennadi2024mri} used SPM12 to create probabilistic atlases from young adult female MRIs, reaching 80\% accuracy, though demographic bias was noted \cite{doraiswamy1992mr,ikram2008pituitary,lamichhane2015age}.

Early innovations by Zukic et al. \cite{zukicetal2011preoperative} and Egger et al. \cite{eggeretal2012pituitary,eggeretal2013segmentation} applied balloon inflation, Grow-cut, and graph-based methods for PA segmentation, achieving scores between 75\% and 81\%, respectively. Sun et al. \cite{sunetal2017random} combined random walk initialization with active contours to enhance segmentation accuracy (88\%). Meanwhile, Thias et al. \cite{thiasetal2019brain} and Kumar et al. \cite{kumaretal2020brain} reported strong results (up to 84\%) using active contour and edge-based approaches on the Cheng dataset \cite{cheng2017brain}. These methods demonstrate that minimal-intervention tools can still yield reliable outcomes, especially for structurally distinct or homogeneous lesions.

\subsection{Comparative Analysis}

Automatic methods, particularly deep learning-based approaches, generally outperform semi-automatic techniques in segmentation accuracy and multi-region coverage (Figure \ref{fig:boxplot_dice_scores_by_type}). U-Net variants have achieved high Dice scores (Figure \ref{fig:pg_pa_unet_vs_others_boxplot}). These methods also support simultaneous segmentation of PA and PG, which is clinically useful when analyzing both lesion and healthy tissue.

A key difference lies in user interaction. Automatic models require no input once trained, enabling consistent, rapid inference. Semi-automatic methods like \textit{GrowCut} \cite{eggeretal2012pituitary}, on the other hand, depend on user input for tasks like boundary marking, which can introduce variability but also enhance precision in challenging cases.

In terms of computational demand, deep learning requires extensive training resources, as highlighted by Wu et al. \cite{wuetal2021deep} and Li et al. \cite{lietal2023preoperatively}. Semi-automatic techniques, such as balloon inflation \cite{zukicetal2011preoperative} and graph-based active contouring \cite{sunetal2017random}, are computationally lighter and suitable for low-resource settings.

Clinically, automated methods suit high-throughput needs, while semi-automatic tools remain relevant where expert guidance is feasible or data are limited. Region-based methods like those by Thias et al. \cite{thiasetal2019brain} and Kumar et al. \cite{kumaretal2020brain} still deliver strong performance for focused tasks. Ultimately, data availability remains crucial to deep learning success.


\subsection{Challenges and Limitations}

Despite promising advances, the studies reviewed exhibit several recurring limitations that impact the interpretability, comparability, and clinical applicability of segmentation models for PG and PA.

One notable challenge is the inconsistent reporting of tumour characteristics—particularly the size of PAs. Many studies, such as Jiang et al. \cite{jiangetal2022improved} and Gab Allah et al. \cite{gaballahetal2023edge}, reported high Dice scores for PA segmentation but did not specify whether the tumors were micro, macro, or giant adenomas. Since larger structures tend to yield higher Dice scores due to clearer anatomical boundaries \cite{yaakub2020brain}, and lower surface-to-volume-ratios, this lack of stratification limits the ability to fairly assess model performance across different tumor complexities. By contrast, studies such as Shu et al. \cite{shuetal2021three} and \textit {Li et al. (2021)} \cite{lietal2021image} provided more granular tumor size classifications, offering a more nuanced understanding of model capabilities across lesion types.

Furthermore, the overwhelming reliance on Dice or Jaccard scores—without complementary metrics like Average Symmetric Surface Distance (ASSD)—restricts insight into boundary accuracy, particularly in smaller or irregularly shaped lesions. Although Dice is widely used, the Jaccard index provides a stricter overlap evaluation: $J = \frac{|A \cap B|}{|A \cup B|}$; $D = \frac{2|A \cap B|}{|A| + |B|}$. These metrics are mathematically related as $D = \frac{2J}{1 + J}$ and $J = \frac{D}{2 - D}$ \cite{jaccard1907distribution,dice1945measures}. Jaccard theoretically controls Dice \cite{bertels2019optimizing} and offers better numerical discrimination for high overlaps.

Data limitations also remain a common concern. While some studies, such as Cerný et al. \cite{cernyetal2023fully}, leveraged relatively large (n=521) internal datasets, most lacked external validation. High-performing models like those in Rai et al. \cite{rai2024two} and Alqhatani et al. \cite{alqhtani2024improved} achieved Dice scores above 95\% using the Cheng 2015 dataset, yet their results were not validated on independent clinical data. On the other end of the spectrum, Wu et al. \cite{wuetal2021deep} demonstrated how segmentation models trained on relatively small datasets (n=155) struggled to perform consistently across architectures, with Dice scores dropping below 40\%. These examples underscore the importance of both data diversity and cross-institutional validation when evaluating generalizability.

Another important limitation is the exclusive reliance on MRI as the imaging modality across all studies. While MRI remains the standard for visualizing pituitary structures, none of the reviewed segmentation pipelines incorporated clinical or biochemical data—such as hormonal assays—that are routinely used for adenoma classification and management. Even studies exploring advanced techniques, such as radiomics and predictive modeling in Li et al. (2023) \cite{lietal2023preoperatively}, remained constrained to imaging-derived features. The absence of multimodal data integration limits the clinical interpretability and potential diagnostic value of the models.

While semi-automatic methods present an alternative that often requires less data and computational power, they are not without limitations. Algorithms such as Grow-cut \cite{eggeretal2012pituitary}, random walk-based models \cite{sunetal2017random}, and balloon inflation techniques \cite{zukicetal2011preoperative} all require varying degrees of user input. Manual initialization introduces inter-operator variability, reducing reproducibility—especially for less experienced users. Additionally, while methods like those proposed by Banday \& Mir \cite{bandayandmir2017enhancement} and Kumar et al. \cite{kumaretal2020brain} showed high Dice scores, their application remains constrained to relatively homogeneous tumor presentations, with less evidence supporting their effectiveness in complex or ambiguous anatomical cases.

Finally, atlas-based segmentation approaches, while historically useful even in small brain structures such as the piriform cortex \cite{steinbart2023automatic}, also face limitations when applied to PG. Isambert et al. \cite{isambertetal2008evaluation} employed a fully automatic pipeline using atlas-based automatic segmentation software (ABAS) - developed from a synthetic MRI \cite{bondiau2005atlas}. While this approach enabled high anatomical consistency during development, the use of artificially generated data may have introduced domain discrepancies when applied to real clinical images—potentially contributing to the relatively low Dice score of 30\% reported for PG segmentation. In contrast, Zennadi et al. \cite{zennadi2024mri} (Dice score of 80\%) used a maximum probability atlas (MPA) within the SPM environment, requiring multi-step normalization and manual parameter tuning—characteristics that justify its classification under semi-automatic methods.

Collectively, these challenges emphasize the need for larger, diverse, and well-annotated datasets; standardized reporting of tumor features; inclusion of multimodal clinical data; and external validation. As segmentation models advance, these methodological improvements will be critical to ensuring their reliability, generalizability, and eventual clinical adoption.


\section{Future Directions}

Future work should focus on developing generalizable models for PG and PA segmentation across diverse imaging protocols and populations, supported by large, multi-institutional annotated datasets. Integrating multimodal inputs—such as clinical or endocrine markers—may enhance diagnostic relevance. Semi-automatic tools should minimize operator input via intuitive design. Lastly, standardized benchmarking and external validation are essential for fair comparisons and clinical translation.

\section{Conclusion}

This review set out to assess the accuracy of automated segmentation algorithms for pituitary adenomas (PA) and the pituitary gland (PG). Deep learning models—particularly U-Net variants—have shown strong performance, with reported Dice scores reaching 96\% for PA and 89\% for PG, indicating high accuracy and clinical promise.

Many studies lacked tumor size stratification or external validation, limiting the generalizability of results. Semi-automatic methods remain relevant where data or computational resources are limited, achieving Dice scores above 80\% in focused applications.

Overall, automated segmentation tools are increasingly accurate and scalable, but their clinical adoption will depend on improvements in dataset diversity, external validation, and integration with clinical workflows. Answering these challenges is essential for reliable, real-world use in diagnosis, treatment planning, and monitoring of pituitary disorders.

\bmhead{Acknowledgements}
We would like to thank Dr.~Ayisha Al-Busaidi, Consultant Neuroradiologist at King’s College Hospital London, for her valuable clinical guidance and expertise throughout this review.

\subsection{Registration and Protocol}

This study is registered with PROSPERO under registration ID CRD42023407127. We adhered as closely as possible to the PRISMA guidelines. We utilized its framework to ensure rigor and transparency in our systematic review process. This included comprehensive search strategies, clearly defined inclusion and exclusion criteria, and a structured approach to data extraction and synthesis.

\subsection{Support}

The primary author is a PhD student funded by the Africa International Postgraduate Research (PGR) Scholarship, provided by the Centre for Doctoral Studies at King's College London. 

The School of Biomedical Engineering and Imaging Sciences is supported by the Wellcome EPSRC Centre for Medical Engineering at King’s College London (WT
203148/Z/16/Z) and the Department of Health via the National Institute for Health Research (NIHR) comprehensive Biomedical Research Centre award to Guy’s \& St Thomas’ NHS Foundation Trust in partnership with King’s College London and King’s College Hospital NHS Foundation Trust. For the purposes of open access, the authors have applied a Creative Commons Attribution (CC BY) licence to any
Accepted Author Manuscript version arising, in accordance with King’s College London’s Rights Retention policy.

\subsection{Competing Interest}

The authors have no competing interests to declare.

\section*{Declarations}

The authors have no conflicts of interest to disclose.

\begin{appendices}




\section{Studies Included by Inclusion Criteria but Excluded Later}\label{secA1}
Table \ref{tab:excluded_studies_1} provides a summary of the studies that were initially included based on the inclusion criteria but were subsequently excluded after a full-text review. The reasons for exclusion are highlighted in the Table.

\begin{table}[h]
\centering
\caption{Summary of Studies Excluded After Full-Text Review}
\label{tab:excluded_studies_1}
\begin{tabular}{p{70mm} p{65mm}} 
\toprule
\textbf{Reason for Exclusion} & \textbf{Study} \\ 
\midrule
No segmentation results for PG or PA provided & \cite{kumarandKumar2023human}, \cite{hashmiandOsman2022brain}, \cite{divyaetal2022enhanced}, \cite{alnaggaretal2022mri}, \cite{balamuruganandgnanamanoharan2021genetic}, \cite{kalametal2021detection}, \cite{pranavandsamhita2021automated}, \cite{guptaetal2021mag}, \cite{isunuriandkakarla2019brain},  \cite{eggeretal2012medical}, \cite{wisaengandsa-ngia2023brain}, \cite{jeribiandperumal2023lesion}, \cite{abdandalsaadi2023automated}, \cite{sahooetal2023efficient}, \cite{singhetal2023detection}, \cite{singhandlobiyal2024comparative}, \cite{krishnasamyandponnusamy2023deep}, \cite{vizzaetal2022image}, \cite{chakiandwozniak2023deep},
\cite{muis2023comparison}, \cite{singhandGautam2023pituitary}, \cite{guptaetal2023brain}, \cite{abd2023automated}, \cite{kordnoori2024deep}, \cite{muhammad2023ensemble},  \cite{kotti2024multi}, \cite{shanthala2023automated}, \cite{krishnan2024multiscale}, \cite{mehta2022computer}, \cite{denes2023brain}, \cite{belaid2024brain}, \cite{roy2024prediction} \cite{kumar2023brainNotIncluded} \cite{rosa2024evaluating}, \cite{kumarNext2023multi}, \cite{geetha2024hybrid}, \cite{waghere2024robust}, \cite{li2023multitask}, \cite{alshomrani2024unified}, \cite{andleeb2023deep}, \cite{jader2024ensemble}, \cite{shirwaikar20243d}, \cite{sindhu2024elevation}, \cite{yan2023deep}, \cite{gargya2024cad}, \cite{almufareh2024automated}, \cite{sindhu2024elevation} \\
\midrule
No overlap metrics provided & \cite{renzetal2011accuracy}, \cite{wongetal2014estimating}, \cite{alhinaietal2011comparison} \\
\midrule
Similar approach with a more detailed study already included & \cite{alqhtani_removed2024contrast}, \cite{asiri_removed2024optimized}, \cite{li2024identificationNotIncluded} \\
\midrule
Book chapter & \cite{abinashetal2022efficient}, \cite{rajagopalandJose2021efficient} \\
\midrule
Segmentation results derived from non-standard Dice and Jaccard formulas & \cite{sunsuhiandJose2022adaptive} \\
\midrule 
Retracted Study & \cite{haqetal2022retracted} \\
\bottomrule 
\end{tabular}
\end{table}

\end{appendices}
\clearpage

\bibliography{sn-bibliography}

\end{document}